\documentclass[12pt]{article}

\usepackage{amsfonts,amssymb,amsmath,amscd}
\usepackage[ps,dvips,matrix,arrow,frame,import,curve,color]{xy}
\newcommand{\ra}{\rightarrow}

\newcommand{\nn}{{\nonumber}}

\newcommand{\ZZ}{{\mathbb Z}}
\newcommand{\RR}{{\mathbb R}}
\newcommand{\CC}{{\mathbb C}}

\newcommand{\bi}{{\bar i}}
\newcommand{\bj}{{\bar j}}
\newcommand{\bk}{{\bar k}}

\newcommand{\tphi}{{\tilde\phi}}

\renewcommand{\part}{\partial}

\newcommand{\bpartial}{{\bar\partial}}

\newcommand{\hc}{{\hat c}}
\newcommand{\bQ}{{\bar Q}}
\newcommand{\trho}{{\tilde\rho}}
\newcommand{\tpsi}{{\tilde\psi}}
\newcommand{\tp}{{\tilde p}}

\newcommand{\bp}{{\bar p}}

\renewcommand{\bi}{{\bar i}}
\renewcommand{\bj}{{\bar j}}
\renewcommand{\bk}{{\bar k}}
\newcommand{\bz}{{\bar z}}
\newcommand{\bw}{{\bar w}}
\newcommand{\bD}{{\bar D}}
\newcommand{\bl}{{\bar l}}

\newcommand{\eps}{\epsilon}
\newcommand{\beps}{{\bar\epsilon}}
\newcommand{\vac}{{|{\rm vac}\rangle}}

\title{Chiral de Rham complex and the half-twisted sigma-model}

\author{Anton Kapustin \\{\small \it California Institute of
Technology, Pasadena, CA 91125, U.S.A.}}

\begin{document}

\begin{titlepage}

\maketitle

\begin{abstract}

On any Calabi-Yau manifold $X$ one can define a certain sheaf of
chiral $N=2$ superconformal field theories, known as the chiral de
Rham complex of $X$. It depends only on the complex structure of
$X$, and its local structure is described by a simple free field theory. 
We show that the cohomology of this sheaf can be identified
with the infinite-volume limit of the half-twisted sigma-model
defined by E. Witten more than a decade ago. We also show that the
correlators of the half-twisted model are independent of the
K\"ahler moduli to all orders in worldsheet perturbation theory,
and that the relation to the chiral de Rham complex can be
violated only by worldsheet instantons.

\end{abstract}

\vspace{-6in}
\parbox{\linewidth}
{\small\hfill \shortstack{CALT-68-2547}} \vspace{6in}

\end{titlepage}

\section{Introduction and summary}

Given a Calabi-Yau manifold $X$ equipped with a K\"ahler class
$\omega$ and a B-field $B\in H^2(X,\RR)$, one can consider the
corresponding supersymmetric sigma-model. This field theory has
$N=(2,2)$ superconformal symmetry, both on the classical and
quantum levels. That is, its Hilbert space carries a unitary
representation of the direct sum of two copies of $N=2$
super-Virasoro algebra, with central charge $\hc=\dim_\CC X$. The
quantum sigma-model is a very complicated object, depending both
on the complexified K\"ahler class $B+i\omega$ and the complex
structure on $X$. In general this field theory can be described
only in the limit when the volume of $X$ is large, using
perturbation theory in the inverse volume.

To obtain a more manageable object, one should ``truncate'' the
sigma-model in some way. In Ref.~\cite{Wittenmir} E. Witten
defined three such truncations, known as the A-model, the B-model,
and the half-twisted model of $X$. The A and B-models are
two-dimensional topological field theories, while the half-twisted
model is a holomorphic $N=2$ superconformal field theory with the
same central charge as the original sigma-model.\footnote{Another
common name for a holomorphic CFT is chiral algebra. In
Ref.~\cite{Kac} they are called conformal vertex algebras.} All
three models are obtained in a uniform way: one starts with the
operator algebra of the $N=2$ sigma-model, shifts the holomorphic
(=left-moving) and antiholomorphic (=right-moving) stress-energy
tensors by derivatives of the corresponding R-currents, and
considers the cohomology of the operator algebra with respect to a
certain nilpotent operator (the BRST charge). The difference
between the models resides in the choice of the shift of the
stress-energy tensors and the choice of the BRST charge. If we
distinguish the generators of the holomorphic and antiholomorphic
$N=2$ super-Virasoro with $\pm$ signs, then the shifts and the
BRST operators are
\begin{multline}\nn
{\rm A-model}: T_-(z)\ra T_-(z)+\frac{1}{2}\partial J_-(z),\
T_+(\bz)\ra T_+(\bz)+\frac{1}{2}\bpartial J_+(\bz), \\
Q_{BRST}=\bQ_+ + Q_- ,\\
{\rm B-model}: \ T_-(z)\ra T_-(z)-\frac{1}{2}\partial J_-(z),\
T_+(\bz)\ra T_+(\bz)+\frac{1}{2}\bpartial J_+(\bz),\\
Q_{BRST}=\bQ_++ \bQ_- ,\\
{\rm half-twisted\ model}:\  T_+(\bz)\ra
T_+(\bz)+\frac{1}{2}\bpartial J_+(\bz), \ Q_{BRST}=\bQ_+.
\end{multline}
Here $Q_\pm$ and $\bQ_\pm$ are Fourier-components of the
corresponding supercurrents; for any field $V(z,\bz)$ the supercommutator
of $\bQ_+$ with $V$ is given by
$$
\{\bQ_+, V\}(z,\bz)=\oint \frac{d\bw}{2\pi i}\,
\bQ_+(\bw)V(z,\bz),
$$
and similarly for other supercharges. Here the integral is over a small contour around $w=z$. In
what follows we choose to redefine the half-twisted model
slightly, by shifting $T_-(z)\ra T_-(z)+\frac{1}{2}\partial
J_-(z)$, as for the A-model. The operator algebra of the
half-twisted model is a holomorphic CFT and contains a copy of the
$N=2$ super-Virasoro algebra. With our slight redefinition, it is
a {\it twisted} $N=2$ super-Virasoro algebra (also known as the
topological $N=2$ algebra).

Witten showed that the B-model is independent of the complexified
K\"ahler class of $X$, while the A-model is independent of the
complex structure of $X$. These topological field theories have
been intensively studied, for a variety of reasons. The
half-twisted model is a much richer object (its state space is
always infinite-dimensional), and has not been much studied. For
general $X$, essentially the only thing we know about the
half-twisted model is its ``Euler characteristic'' defined as
$$
\chi(q,\gamma)={\rm Tr} (-1)^F q^{L_0-\frac{\hc}{8}} \exp(i\gamma J_-),
$$
where $F$ is the total fermion number operator, and
$$
L_0=\oint \frac{dz}{2\pi i}\, z T_-(z),\quad J_-=\oint
\frac{dz}{2\pi i} J_-(z).
$$
The Euler
characteristic is known to coincide with a certain two-variable
elliptic genus of $X$~\cite{Wittenloop}.

A few years ago mathematicians Malikov, Schechtman, and Vaintrob
have defined, for any complex manifold $X$, a sheaf of holomorphic
operator algebras, called the chiral de Rham complex~\cite{MSV}.
Locally, the space of sections of this sheaf is very simple: it is
the tensor product of $n$ copies of the familiar $\beta\gamma-bc$
system, where $n=\dim_\CC X$. It was shown in Ref.~\cite{MSV} that
if one performs arbitrary analytic reparametrizations of the
$\gamma$-fields, there exist analytic transformation laws for the
remaining fields which preserve the Operator Product Expansion
(OPE). This surprising fact allows one to treat the constant modes
of the $\gamma$-fields as local coordinates on $X$, and glue the
locally-defined operator algebras into a sheaf of operator
algebras on $X$. In the case when $X$ is a Calabi-Yau, the authors
of Ref.~\cite{MSV} showed that there is an extra structure on this
sheaf: it is a sheaf of holomorphic $N=2$ superconformal field
theories, whose central charge is equal to $\dim_\CC X$. (Locally,
this is obvious, because the $\beta\gamma-bc$ system has $N=2$
superconformal symmetry. However, it is nontrivial to check that
the generators of the $N=2$ super-Virasoro algebra glue properly
on double overlaps.) The cohomology of this sheaf is a certain
holomorphic $N=2$ SCFT, which is fairly complicated, in general.
As explained in Ref.~\cite{BL}, a suitably defined Euler
characteristic of this sheaf is equal to the elliptic genus of
$X$. This suggests a relation with the half-twisted model of E.
Witten. There are two puzzles which must be resolved to establish
a precise connection. First, physicists usually do not think in
terms of sheaves of CFTs: the sigma-model is a global object, from
the target-space viewpoint. Second, the half-twisted model depends
both on complex and K\"ahler structures of $X$, while the chiral
de Rham complex depends only on the complex structure.

In this note we show that the cohomology of the chiral de Rham
complex can be identified with the infinite-volume limit of the
half-twisted sigma-model. This resolves the second puzzle. As for
the first puzzle, we recall that one can compute sheaf cohomology
in a number of ways. The definition used in Ref.~\cite{MSV}
naturally leads one to the \v{C}ech resolution. On the other hand,
one can also construct a ``Dolbeault resolution'' for the chiral
de Rham complex, whose graded components are soft sheaves of
holomorphic $N=2$ SCFTs.  Thus the sheaf cohomology can be
computed by taking the global sections of the Dolbeault resolution
and computing the cohomology of the Dolbeault differential on the
resulting graded vector space. We show that this recipe for
computing sheaf cohomology  is the infinite-volume limit of the
physical definition of the half-twisted sigma-model. This resolves
the first puzzle.

Having established that the infinite-volume limit of the
half-twisted model is related to the chiral de Rham complex, we
then ask how this result is modified at a large but finite volume.
We show that the half-twisted model does not depend on K\"ahler
moduli to all orders in the large-volume expansion. Thus the
relation between the chiral de Rham complex and the half-twisted
model is perturbatively exact. We find it rather surprising that
such a seemingly complicated object as the half-twisted
sigma-model for an arbitrary Calabi-Yau manifold can be computed
to all orders in perturbation theory by gluing together free
theories ($\beta\gamma-bc$ systems). On the other hand, we expect
that nonperturbative contributions from worldsheet-instanton lead
to K\"ahler moduli dependence, much like in the A-model. Unlike
the A-model, however, the half-twisted model also depends on the
complex structure of $X$.

The content of the paper is as follows. In Section~\ref{cdr} we
describe the chiral de Rham complex and its Dolbeault resolution.
In Section~\ref{htm} we show that the infinite-volume limit of the
half-twisted sigma-model coincides with the cohomology of the
Dolbeault differential acting on the space of global sections of
the Dolbeault resolution. In Section~\ref{kahler} we study the
dependence of the half-twisted model on the K\"ahler moduli.
In Section~\ref{concl} we briefly discuss possible generalizations.

In Section~\ref{cdr} we make use of some basic notions of sheaf
theory. For an introduction, see e.g. Chapters 1 and 2 of
Ref.~\cite{Bredon}. Readers with a sheaf-intolerance are advised
to skim Section~\ref{cdr}; the rest of the paper does not use
sheaves.

\section{Chiral de Rham complex and its Dolbeault
resolution}\label{cdr}

Following Ref.~\cite{MSV}, we first describe the chiral de Rham
complex of an affine space $\CC^n$. It is a holomorphic $N=2$ SCFT
which is a tensor product of $n$ copies of the $\beta\gamma-bc$
system. That is, we have free bosonic fields $\phi^i(z), p_j(z)$
and free fermionic fields $\psi^i(z),\rho_i(z)$ with the
nontrivial OPE
$$
\phi^i(z)p_j(w)\sim \frac{\delta^i_j}{z-w},\quad
\rho_i(z)\psi^j(w)\sim \frac{\delta^j_i}{z-w}.
$$
(Our convention for the fields $p_i$ differs by a sign from that
of Ref.~\cite{MSV}. This difference in conventions accounts for
some extra minus signs below.) Such OPEs follow from quantizing a
quadratic action
\begin{equation}\label{act1}
S=\frac{1}{\pi}\int d^2z \left(p_i\bpartial \phi^i+\rho_i\bpartial
\psi^i\right).
\end{equation}
The equations of motion following from this action imply that the
fields can be expanded into Laurent series:
\begin{equation}\nn
\begin{array}{ll}
\phi^i=\sum_{n\in\ZZ} \phi^i_n z^{-n},&
p_i=\sum_{n\in\ZZ} p_{i,n} z^{-n-1},\\
\psi^i=\sum_{n\in\ZZ}\psi^i_n z^{-n},& \rho_i=\sum_{n\in\ZZ}
\rho_{i,n} z^{-n-1},
\end{array}
\end{equation}
{}From the OPEs we get the commutation relations:
$$
[\phi^i_n,p_{j,m}]=\delta^i_j \delta_{n,-m},\quad
\{\psi^i_n,\rho_{j,m}\}=\delta^i_j \delta_{n,-m},
$$
with all other supercommutators vanishing. The state space of the
theory is the tensor product of the space of holomorphic functions
of variables $\phi^i_0,\psi^i_0$ and the Fock space of the
``oscillator'' modes (the ones with $n\neq 0$). The Fock space contains
a vacuum vector $\vac$ satisfying
$$
\phi^i_n\vac=p_{i,n}\vac=\psi^i_n\vac=\rho_{i,n}\vac=0,\quad
\forall n>0,
$$
and is generated by acting on the vacuum vector with $n<0$
oscillators. The remaining zero-modes, $p_{i,0}$ and $\rho_{i,0}$,
act as differential operators on holomorphic functions of
$\phi^i_0$ and $\psi^i_0$:
$$
p_{i,0}=-\frac{\partial}{\partial\phi^i_0},\quad
\rho_{i,0}=\frac{\partial}{\partial\psi^i_0}.
$$

The stress-energy tensor is defined to be
\begin{equation}\label{T}\nn
T_-(z)=-p_i(z)\partial \phi^i(z)-\rho_i(z) \partial\psi^i(z).
\end{equation}
(Here and below we omit the symbol of normal-ordering). The
supercurrents are
\begin{equation}\label{super}\nn
Q_-(z)=-\psi^i(z) p_i(z),\quad \bQ_-(z)=\rho_i(z)
\partial\phi^i(z).
\end{equation}
The R-current is
\begin{equation}\label{J}\nn
 J_-(z)=\rho_i(z)\psi^i(z).
\end{equation}
Computing the OPE of these currents, one finds that they (or
rather the coefficients of their Laurent expansions) form a
topologically twisted $N=2$ super-Virasoro algebra with central
charge $\hc=n$. The supersymmetry transformations of the fields
are computed from the OPE:
$$
\{Q_-,V(w)\}=\oint \frac{dz}{2\pi i}\, Q_-(z)V(w),\quad
\{\bQ_-,V(w)\}=\oint \frac{dz}{2\pi i}\, \bQ_-(z)V(w).
$$
In this way one finds:
\begin{equation}\label{cdrSUSY}
\begin{array}{ll}
\{Q_-,\phi^i\}=\psi^i, & \{\bQ_-,\phi^i\}=0,\\
\{Q_-,p_i\}=0, & \{\bQ_-,p_i\}=-\partial\rho_i,\\
\{Q_-,\psi^i\}=0, & \{\bQ_-,\psi^i\}=\partial\phi^i,\\
\{Q_-,\rho_i\}=-p_i, & \{\bQ_-,\rho_i\}=0.
\end{array}
\end{equation}
{}From the worldsheet viewpoint, $\rho_i$ and $p_i$ are 1-forms,
while $\psi^i$ and $\phi^i$ are 0-forms. One can untwist the
theory by subtracting $\frac{1}{2}\partial J_-$ from the
stress-energy tensor; then $\rho_i$ and $\psi^i$ become worldsheet
spinors of the same chirality.

Ordinarily, in free field theory one considers composite fields
which are normal-ordered polynomials of the fundamental fields
(which in this case are $\phi^i,p_i,\psi^i,\rho_i$) and their
derivatives. In the present case, since the OPE of $\phi^i$ with
itself is trivial, it makes sense to consider composite operators
where arbitrary analytic functions of $\phi^i$ (and polynomial
functions of other fields and derivatives of $\phi^i$) are
allowed. This is explained in detail in section 3.1 of
Ref.~\cite{MSV}. Briefly speaking, given a function $f(\phi)$, one
replaces the variable $\phi$ with the corresponding Laurent series
$\phi(z)$ and reexpands in powers of $z$. Although the coefficient
of each power of $z$ is an infinite sum of operators on the tensor
product of the bosonic Fock space and the space of holomorphic
functions, one can show that it is well-defined.

To define a sheaf of holomorphic $N=2$ SCFTs, one has to invent a
transformation law for the fields which is compatible with the
postulated OPEs and preserves the expressions for the
stress-energy tensor, supercurrents, and the R-current. Given a
change of coordinates
$$
\tphi^i=g^i(\phi),\quad \phi^i=f^i(\tphi),
$$
an obvious transformation law for the fermions is
$$
\tpsi^i=\frac{\partial g^i}{\partial \phi^j} \psi^j,\quad
\trho_i=\frac{\partial f^j}{\partial \tphi^i} \rho_j.
$$
The correct transformation law for the bosonic field $p_i$ is less
obvious:
\begin{equation}\label{inhomo}
\tp_i=\frac{\partial f^j}{\partial \tphi^i} p_j+\frac{\partial^2
f^k}{\partial\tphi^i\partial\tphi^l}\frac{\partial g^l}{\partial
\phi^r}\rho_k\psi^r.
\end{equation}
In Ref.~\cite{MSV} this transformation law was motivated by its
compatibility with the OPE (Th. 3.7 of Ref.~\cite{MSV}). We will
see in the next section how this transformation law arises
naturally in the half-twisted model.

Finally, one has to check that the $N=2$ super-Virasoro algebra is
preserved under coordinate change. It turns out that this is true
if and only if all Jacobians
$$
\det \frac{\partial g^i}{\partial \phi^j}
$$
are  constant. This is possible to achieve if and only if
$c_1(X)=0$, i.e. if $X$ is a Calabi-Yau
manifold.\footnote{Strictly speaking, this condition is equivalent
to the Calabi-Yau condition only if $X$ is simply-connected. If
$H^1(X)\neq 0$, then it is possible that $c_1(X)=0$, but the
canonical line bundle of $X$ is nontrivial, as a holomorphic line
bundle. {}From the physical viewpoint, this means that the BRST
cohomology does not contain the spectral flow operator.} Thus for
any Calabi-Yau $X$ one obtains a sheaf of holomorphic $N=2$ SCFTs.
By definition, this is the chiral de Rham complex of $X$, denoted
$\Omega_X^{ch}$. The space of sections of this sheaf over an open
set $U$ will be denoted $\Omega_X^{ch}(U)$.

We are interested in the sheaf cohomology of the chiral de Rham
complex, which, on general grounds, is a holomorphic $N=2$ SCFT.
The degree-0 part is simply the space of global sections of the
chiral de Rham complex, while higher-degree cohomology can be defined using
either the \v{C}ech approach or any soft resolution. In order to
make a connection with the half-twisted model, we would like to
consider a Dolbeault-like resolution. Locally, this is achieved by
introducing bosonic variables $\phi^\bi$ complex--conjugate to the
zero-modes of $\phi^i$ and fermionic (i.e. odd) variables
$\psi^\bi$, and enlarging the chiral de Rham complex by allowing
arbitrary smooth functions of $\phi^i$ and $\phi^\bi$ and
polynomial functions of the remaining fields, their
$\partial$-derivatives, and the variables $\psi^\bi$. Note that
while $\phi^i$ are fields (i.e. they are power series in $z$ and
$\bz$), $\phi^\bi$ are simply complex variables. Thus it is not
completely obvious that the procedure is well-defined. Recall that
we started with polynomial functions of the fields $\phi^i$, but
then replaced them with arbitrary holomorphic functions. As
explained in Ref.~\cite{MSV}, this construction has the following
generalization: instead of holomorphic function one can take an
arbitrary supercommutative algebra which contains the
polynomial algebra of $\phi^i$ as a subalgebra and is equipped
with an action of the Lie algebra of polynomial vector fields
$\frac{\partial}{\partial \phi^i}$ which satisfies the Leibniz
rule. We are dealing with a special case, where the algebra in
question is the algebra of smooth forms of type $(0,p)$ (for some
$p$).

In this way we get a new sheaf of holomorphic $N=2$ SCFTs graded
by the $\psi^\bi$-degree. We will denote it $\Omega_X^{ch,Dol}$. The
Dolbeault operator
$$
d_{Dol}=\psi^\bi\frac{\partial}{\partial \phi^\bi}
$$
makes $\Omega_X^{ch,Dol}$ into a complex of sheaves. It is easy to
see that this is a resolution of the chiral de Rham complex: this
follows from the usual $\bpartial$-Poincar\'{e} lemma and the fact
that  $d_{Dol}$ commutes with the normal-ordering of operator
product. We will call this complex of sheaves the Dolbeault
resolution of the chiral de Rham complex.

It is also easy to see that the Dolbeault resolution is soft. The
proof is the same as for the ordinary Dolbeault resolution and
makes use of the existence of partition of unity for smooth
functions. Softness implies that we can compute the sheaf
cohomology of the chiral de Rham complex by considering the space
of global sections of the Dolbeault resolution
$\Omega_X^{ch,Dol}(X)$, and then computing the cohomology of
$d_{Dol}$. As we will see in the next section, this global
approach to computing the sheaf cohomology is much better suited
for comparison with the half-twisted sigma-model.

\section{The half-twisted model and its large volume
limit}\label{htm}

Now we turn to the half-twisted sigma-model whose target is a
Calabi-Yau manifold. Recall that we decided to twist both the
holomorphic and antiholomorphic stress-energy tensors, so the
action is the same as for the A-model~\cite{Wittenmir}:
\begin{multline}\label{act2}
S=\frac{1}{\pi}\int d^2z
\left[\frac{1}{2}\left(g_{i\bj}+B_{i\bj}\right)
\partial\phi^i\bpartial\phi^\bj+\frac{1}{2}\left(g_{i\bj}-B_{i\bj}\right)\bpartial\phi^i\partial\phi^\bj+
\rho_i \bD\psi^i\right.\\
\left. +\rho_\bi D\psi^\bi+R^{i\bi}_{\ \ j\bj}\rho_i\rho_\bi
\psi^j\psi^\bj\right].
\end{multline}
Here $\phi^i(z,\bz)$ is a coordinate representation of a map from the worldsheet $\Sigma$ to the target $X$,
$\psi^i,\psi^\bi$ are fermionic fields taking values in $\phi^* TX^{1,0}$ and $\phi^*TX^{0,1}$,
respectively, and $\rho_i,\rho_\bi$ are fermionic fields taking values in 
$K_\Sigma\otimes \phi^* T^*X^{1,0}$ and ${\bar K}_\Sigma\otimes\phi^*T^*X^{0,1}$,
respectively, where $K_\Sigma$ is the canonical line bundle of $\Sigma$. Further, $g_{i\bj}(\phi)$
is the K\"ahler metric on $X$, $R^{i\bi}_{\,\, j\bj}$ is the corresponding Riemann tensor, $B_{i\bj}(\phi)$ is the B-field (which is assumed here to be closed and of type $(1,1)$),
and $D$ and $\bD$ are holomorphic and antiholomorphic covariant
derivatives on the vector bundle $\phi^* TX$:
$$
\bD\psi^i=\bpartial\psi^i+\Gamma^i_{jk}\left(\bpartial
\phi^j\right) \psi^k,\quad
D\psi^\bi=\partial\psi^\bi+\Gamma^\bi_{\bj\bk}
\left(\partial\phi^\bj\right)\psi^\bk.
$$
We are going to consider the infinite-volume limit
$g_{i\bj}\ra\infty$. In this limit nonperturbative effects, such
as worldsheet instantons, are irrelevant, and the B-field does not affect the dynamics. If we
choose a purely imaginary B-field, $B=i\omega$, where $\omega$ is
the K\"ahler form, then the first term in Eq.~(\ref{act2})
vanishes, and the action becomes
\begin{equation}\label{act3}
S=\frac{1}{\pi}\int d^2z \left[
g_{i\bj}\bpartial\phi^i\partial\phi^\bj+ \rho_i \bD\psi^i
+\rho_\bi D\psi^\bi+R^{i\bi}_{\ \ j\bj}\rho_i\rho_\bi
\psi^j\psi^\bj\right].
\end{equation}
Alternatively, one can simply set $B=0$ and note that the
first term in Eq.~(\ref{act2}) differs from the second one by
$\frac{1}{4\pi}\int \phi^*\omega$, which vanishes for
topologically trivial string worldsheets. Therefore, if one neglects the
contributions of worldsheet instantons, one can replace Eq.~(\ref{act2}) with Eq.~(\ref{act3}).

Consider now the following action:
\begin{multline}\label{act4}
S^{(1)}=\frac{1}{\pi}\int d^2z \left[ p_i\bpartial\phi^i+
p_\bi\partial\phi^\bi+
\rho_i\bpartial\psi^i+\rho_\bi\partial\psi^\bi\right.\\
\left. -g^{\bj
i}\left(p_i-\Gamma^k_{il}\rho_k\psi_l\right)\left(p_\bj-\Gamma^\bk_{\bj\bl}\rho_\bk\psi^\bl\right)
+R^{i\bi}_{\ \ j\bj}\rho_i\rho_\bi \psi^j\psi^\bj\right].
\end{multline}
The equations of motion for the bosonic fields $p_i$ and $\bp_\bi$
are algebraic and read
\begin{equation}\label{psolved}
p_i=g_{i\bj}\partial\phi^\bj+\Gamma^j_{ik}\rho_j\psi^k,\quad
p_\bj=g_{i\bj}\bpartial\phi^i+\Gamma^\bi_{\bj\bk}\rho_\bi\psi^\bk.
\end{equation}
Substituting the expressions for $p_i,p_\bj$ back into
Eq.~(\ref{act4}), one finds the action Eq.~(\ref{act3}). Thus
Eq.~(\ref{act3}) and Eq.~(\ref{act4}) define equivalent theories.
The action Eq.~(\ref{act4}) is more convenient for taking the
infinite-volume limit. In this limit the inverse metric $g^{\bj i}$
goes to zero, so it is clear that we should keep the fields
$\phi^i,p_i,\psi_i,\rho_i$ and their complex conjugates fixed. The
limiting action is quadratic:
\begin{equation}\label{act5}
S^{(1)}_\infty=\frac{1}{\pi}\int d^2z \left[p_i\bpartial\phi^i+
p_\bi\partial\phi^\bi+
\rho_i\bpartial\psi^i+\rho_\bi\partial\psi^\bi\right].
\end{equation}
Quantization of this action gives a free $N=2$ SCFT. This theory
is essentially a tensor product of the theory defined by the
action Eq.~(\ref{act1}) and its complex-conjugate. This statement
would be exactly true if we allowed only operators which depend
polynomially on $\phi^i$ and $\phi^\bi$. But since the zero-modes
of $\phi^i(z)$ and $\phi^\bi(\bz)$ are local coordinates on $X$,
one should allow arbitrary smooth functions of $\phi^i$ and
$\phi^\bi$. (The dependence on the rest of the fields and the
derivatives of all the fields is polynomial.)

On the other hand, the limiting $N=2$ SCFT is also similar to the
space $\Omega_X^{ch,Dol}(X)$. The difference is that in the latter
case only the zero modes of the antiholomorphic fields
$\phi^\bi(\bz),\psi^\bi(\bz)$ are retained. Let us show that this difference
becomes irrelevant if we compare the $\bQ_+$ cohomology of the
theory Eq.~(\ref{act5}) with the $d_{Dol}$ cohomology of
$\Omega_X^{ch,Dol}(X)$.

The BRST transformations of the half-twisted model (before taking
the infinite-volume limit) read~\cite{Wittenmir}:
\begin{align}\label{BRSTvar}
\delta\phi^i=0, &\ \ \ \ \delta\phi^\bi=\eps\psi^\bi,\nn\\
\delta\psi^i=0, &\ \ \ \ \delta\psi^\bi=0,\\
\delta\rho_i=0, &\ \ \ \ \delta\rho_\bi=-\eps
g_{i\bi}\bpartial\phi^i-\eps\Gamma^\bk_{\bi\bj}\rho_\bk\psi^\bj=-\eps
p_\bi.\nn
\end{align}
Here $\eps$ is an odd parameter of the BRST transformation. The
BRST variation of $p_\bi$ vanishes identically,
while the BRST variation of $p_i$ vanishes upon using the
equations of motion. This can be shown using Eq.~(\ref{BRSTvar})
and Eq.~(\ref{psolved}). The operator $\bQ_+$ generating these
transformations has the following simple form:
$$
\bQ_+=-\int d\bz\, \psi^\bi p_\bi .
$$
In the infinite-volume limit, we can expand the fields with
antiholomorphic indices into antiholomorphic Laurent series:
\begin{equation}\nn
\begin{array}{ll}
\phi^\bi=\sum_{n\in\ZZ} \phi^\bi_n \bz^{-n},&
p_\bi=\sum_{n\in\ZZ} p_{\bi,n} \bz^{-n-1},\\
\psi^\bi=\sum_{n\in\ZZ}\psi^\bi_n \bz^{-n},&
\rho_\bi=\sum_{n\in\ZZ} \rho_{\bi,n} \bz^{-n-1},
\end{array}
\end{equation}
whose coefficients are operators satisfying the commutation
relations
$$
[\phi^\bi_n,p_{\bj,m}]=\delta^\bi_\bj \delta_{n,-m},\quad
\{\psi^\bi_n,\rho_{\bj,m}\}=\delta^\bi_\bj \delta_{n,-m}.
$$
The modes with $n\neq 0$ (``oscillators'') are represented by
operators in the Fock space in the standard manner by postulating
$$
\phi^\bi_n\vac
=\psi^\bi_n\vac=p_{\bi,n}\vac=\rho_{\bi,n}\vac=0,\quad \forall
n>0.
$$
In terms of the coefficients of the Laurent expansion, the
formula for $\bQ_+$ reads:
$$
\bQ_+=-\sum_{n\in\ZZ}\psi^\bi_{-n}p_{\bi,n}.
$$
To compute the cohomology of $\bQ_+$, we note that the space of
states is a tensor product of the space of zero-modes and the
oscillator modes (modes with $n\neq 0$). The operator $\bQ_+$ is a
sum of terms, each of which acts only in one factor and does not
mix modes with different $n$. For $n\neq 0$, it is easy to see
that the cohomology of $\bQ_+$ in the Fock space generated by
$\phi^\bi_{-n},p_{\bi,-n},\psi^\bi_{-n},\rho_{\bi,-n}$ is
one-dimensional. Indeed, from the physical viewpoint the problem is identical
to that of two independent supersymmetric harmonic oscillators, with the supersymmetry
generator playing the role of $\bQ_+$. The cohomology of this system is well-known to
be one-dimensional. {}From the mathematical viewpoint, we are dealing with the tensor
product of a Koszul complex, corresponding to a free resolution of a point on $\CC^n$,
and the algebraic de Rham complex of $\CC^n$.

Thus we may as well drop all the right-moving modes with $n\neq
0$. We are left with the theory where the only antiholomorphic
modes are the zero-modes
$\phi^\bi_0,p_{\bi,0},\psi^\bi_0,\rho_{\bi,0}$ satisfying the
canonical commutation relations
$$
[\phi^\bi_0,p_{\bj,0}]=\delta^\bi_\bj,\quad
\{\psi^\bi_0,\rho_{\bj,0}\}=\delta^\bi_\bj.
$$
These are quantized in the standard manner: the states are
arbitrary functions of the variables $\phi^\bi_0$ and
$\psi^\bi_0$, while the canonically-conjugate variables are
realized by differential operators:
$$
p_{\bi,0}=-\frac{\partial}{\partial\phi^\bi_0},\quad
\rho_{\bi,0}=\frac{\partial}{\partial\psi^\bi_0}.
$$
The operator $\bQ_+$ reduces to
$$
\bQ_+=\psi^\bi_0 \frac{\partial}{\partial\phi^\bi_0}.
$$
The space of states of this SCFT is exactly
$\Omega_X^{ch,Dol}(X)$. This is obvious locally; we discuss the
transformation law for going from chart to chart below. The
operator $\bQ_+$ acts on the space of states of this SCFT as
$d_{Dol}$. Thus the $\bQ_+$-cohomology of the sigma-model is
isomorphic to the $d_{Dol}$-cohomology of the space
$\Omega_X^{ch,Dol}(X)$. The latter is the same as the sheaf
cohomology of the chiral de Rham complex.

Further, one can now see where the peculiar transformation law for
$p_i$ comes from: while $\partial\phi^\bj$ transforms
homogeneously, the expression for $p_i$ also involves the term
$$
\Gamma^j_{ik}\rho_j\psi^k.
$$
The well-known transformation law for the Christoffel symbols
$$
\tilde\Gamma^i_{jk}=\frac{\partial\tphi^i}{\partial\phi^l}\frac{\partial\phi^m}{\partial\tphi_j}
\frac{\partial\phi^n}{\partial\tphi_k}\Gamma^l_{mn}+\frac{\partial^2\phi^l}{\partial
\tphi^j\partial\tphi^k}\frac{\partial\tphi^i}{\partial\phi^l}.
$$
then implies the transformation law Eq.~(\ref{inhomo}). The fields $\rho_i$ and $\psi^i$
transform homogeneously, as expected.

Finally, let us consider the supersymmetry transformations
generated by $Q_-$ and $\bQ_-$. If we denote the corresponding odd
parameters by $\eps$ and $\beps$, then according to
Ref.~\cite{Wittenmir}, they are given by
\begin{equation}\nn
\begin{array}{ll}
\delta\phi^i&=\eps \psi^i,\\
\delta\phi^\bi&=-\beps g^{\bi i}\rho_i,\\
\delta\psi^i&=\beps\partial\phi^i,\\
\delta\psi^\bi&=\beps \Gamma^\bi_{\bj\bk} g^{\bj i}\rho_i\psi^\bk,\\
\delta\rho_i&=-\eps\left(g_{i\bi}\partial\phi^\bi+\Gamma^k_{ij}\rho_k\psi^j\right)=-\eps p_i,\\
\delta\rho_\bi&=-\beps\Gamma^\bj_{\bi\bk}g^{\bk i}\rho_i\rho_\bj.
\end{array}
\end{equation}
In the infinite-volume limit they become
\begin{align}
\delta\phi^i=\eps\psi^i,&\ \ \ \ \delta\phi^\bi=0,\nn\\
\delta\psi^i=\beps\partial\phi^i,&\ \ \ \ \delta\psi^\bi=0,\nn\\
\delta\rho_i=-\eps p_i,&\ \ \ \ \delta\rho_\bi=0.\nn
\end{align}
We see that in this limit all the antiholomorphic fields are
invariant with respect to all left-moving supersymmetry
transformations. On the other hand, the transformations of the
fields $\phi^i,\psi^i,\rho_i$ agree with Eq.~(\ref{cdrSUSY}). The
transformation law for $p_i$ follows from the above
transformations and properties of the supersymmetry algebra:
\begin{align}
\{ Q_-,p_i\}&=-\{Q_-,\{Q_-,\rho_i\}\}=0,\nn\\
\{\bQ_-,p_i\}&=-\{\bQ_-,\{Q_-,\rho_i\}\}=-\partial\rho_i.\nn
\end{align}
This concludes the demonstration that the infinite-volume limit of
the half-twisted model is isomorphic, as an $N=2$ SCFT, to the
sheaf cohomology of the chiral de Rham complex.

\section{The dependence of the half-twisted model on K\"ahler moduli}\label{kahler}

We will show in this section that to any order in the perturbative
(large-volume) expansion the half-twisted model is independent of
the K\"ahler moduli, but may receive contributions from worldsheet
instantons. Thus the correspondence with the chiral de Rham
complex persists to all orders in perturbation theory, but not
nonperturbatively.

To this end we have to rewrite the action of the half-twisted
sigma-model as a sum of a topological term, a BRST-trivial term,
and the remainder, so that all the K\"ahler-moduli dependence is
in the first two terms. This is very similar to the argument for
the A-model in Ref.~\cite{Wittenmir}.

The BRST transformations for the fields are displayed in
Eq.~(\ref{BRSTvar}). Let
$$
V=-\rho_\bi\partial\phi^\bi-g^{\bl
k}\Gamma^i_{jk}\rho_i\psi^j\rho_\bl.
$$
It is straightforward to check that the action Eq.~(\ref{act2})
can be written in the following way:
$$
S=\frac{1}{4\pi}\int \phi^*(\omega+i B)+\frac{1}{\pi}\int d^2
z\left(\rho_i\bpartial\phi^i+\{\bQ_+,V\}\right).
$$
The first term depends on $\omega$ and $B$, but it contributes
only for topologically nontrivial worldsheets. The second term
contains a BRST-trivial piece and a piece which depends only on
the complex structure of $X$. We conclude that the dependence on
the K\"ahler moduli is absent to all orders in perturbation
theory, if we consider only correlators of BRST-closed operators
(i.e. if we work with the half-twisted sigma-model).

To go beyond perturbation theory, one has to include worldsheet
instantons. As usual, the path-integral localizes on classical
configurations which are BRST-invariant. In the present case, the
condition is that the variation of $\rho_\bi$ must vanish. This
implies that the relevant classical configurations are solutions
of
$$
\bpartial\phi^i=0,
$$
i.e. holomorphic maps from the worldsheet to $X$, just like in the
A-model. It would be of some interest to determine the precise
form of the instanton corrections to the chiral de Rham complex.
This would provide a ``chiral'' generalization of the quantum
cohomology ring.

\section{Concluding remarks}\label{concl}

The definition of the chiral de Rham complex makes sense even if $c_1(X)\neq 0$.
However, in this case one cannot globally define the generators $J_-(z)$ and $Q_-(z)$~\cite{MSV}.
Thus one gets a sheaf of holomorphic operator algebras, but not a sheaf of $N=2$ superconformal
field theories. Note that although the twisted stress-energy tensor $T_-(z)$ can still be defined,
we cannot ``untwist'' it into a ``physical'' stress-energy tensor with a nonzero central charge,
because we lack $J_-(z)$. This is related to the fact that for $c_1(X)\neq 0$ the sigma-model
is not conformally-invariant. 

{}From the physical viewpoint, for $c_1(X)\neq 0$ one should consider both the $\bQ_+$-cohomology
of operators and the $\bQ_+$-cohomology of states in the Ramond sector. While they
are isomorphic vector spaces in the conformal case, this is not so in general. Rather, one
can only say that the latter is a module over the former. If the theory has a mass gap, the cohomology
of states is much smaller than the cohomology of operators (usually, finite-dimensional).

Another very interesting generalization is to replace the fermions $\psi^i$ and $\rho_i$ taking values
in $TX$ and $T^*X$ with fermions taking values in an arbitrary holomorphic vector bundle $E$ and its dual.
This generalization was considered in Refs.~\cite{MSV,GMS}, where it was found that such conformal
field theories can be glued into a sheaf of conformal field theories only if the topological condition
$$
ch_2(E)-ch_2(TX)=0
$$
is satisfied. Here $ch_2(E)$ is the degree-4 part of the Chern character of $E$.
{}From the physical viewpoint, this generalization corresponds to studying the twisted
version of the heterotic sigma-model with $(0,2)$ supersymmetry. The above topological condition is 
the familiar condition of cancellation of worldsheet anomalies for heterotic 
sigma-models.

\section*{Acknowledgments}
A.K. would like to thank Fyodor Malikov and Vassily Gorbounov for
discussions. This work was supported in part by the DOE grant
DE-FG03-92-ER40701.

\end{document}